\documentclass[final,3p,times,twocolumn]{elsarticle}
\usepackage{longtable}

\usepackage{graphics}
\usepackage{amsmath}
\usepackage{amssymb}
\usepackage{natbib}

\usepackage{lineno}
\usepackage{tabularx} 
\usepackage{booktabs}  
\usepackage{multirow}
\usepackage{array}
\usepackage{microtype}
\usepackage{siunitx}
\usepackage{threeparttable} 
\usepackage{graphicx}

\journal{NIM A}

\usepackage{multicol}

\begin{document}

\begin{frontmatter}



\title{A cryogenic gas target for high-intensity radioactive ion beam production at HIRFL-RIBLL}


\author[label1]{Xiao Fang\corref{cor1}}
\ead{fangx26@mail.sysu.edu.cn}
\author[label2,label3]{Shiwei Xu\corref{cor1}}
\ead{shwxu@impcas.ac.cn}
\author[label2,label3]{Longhui Ru}
\author[label1]{Ruiqi Chen}
\author[label2,label3]{Bingshui Gao}
\author[label2,label3]{Song Guo}
\author[label2,label3]{Jun Hu}
\author[label4]{Huiming Jia}
\author[label2,label3]{Xinyue Li}
\author[label4]{Chengjian Lin}
\author[label2,label3]{Enqiang Liu}
\author[label2,label3]{Chengui Lu}
\author[label2,label3]{Junbing Ma}
\author[label5]{Jun Su}
\author[label2,label3]{Xiaodong Tang}
\author[label6]{Jiansong Wang}
\author[label4]{Shengquan Yan}
\author[label4]{Lei Yang}
\author[label1]{Ruojun Yang}
\author[label2,label3]{Yanyun Yang}
\author[label7]{Gaolong Zhang}
\author[label5]{Liyong Zhang}
\author[label2,label3]{Ningtao Zhang}
\author[label2,label3]{Zhichao Zhang}

\cortext[cor1]{Corresponding authors.}

\address[label1]{Sino-French Institute of Nuclear Engineering and Technology, Sun Yat-sen University, Zhuhai, Guangdong 519082, China}
\address[label2]{Institute of Modern Physics, Chinese Academy of Sciences, Lanzhou, Gansu 730000,  China}
\address[label3]{School of Nuclear Science and Technology, University of Chinese Academy of Sciences, Beijing 100049, China}
\address[label4]{China Institute of Atomic Energy, Beijing 102413, China}
\address[label5]{Key Laboratory of Beam Technology of Ministry of Education, School of Physics and Astronomy, Beijing Normal University, Beijing 100875, China}
\address[label6]{School of Science, Huzhou Normal University, Huzhou, Zhejiang 313000, China}
\address[label7]{School of Physics, Beihang University, Beijing 100191, China}

\begin{abstract}
A liquid-nitrogen-cooled cryogenic gas target system has been developed and installed for radioactive ion beam (RIB) production at the Radioactive Ion Beam Line in Lanzhou (RIBLL). Light-element gases ($\mathrm{H}_2$, $\mathrm{D}_2$, and $^4\mathrm{He}$) filled in the target cell were cooled to cryogenic temperatures, with the gas-cell outlet temperature typically monitored at 82--86 K during beam irradiation and operating pressures up to 1000 mbar. The system was used to produce $^{7}\mathrm{Be}$, $^{16}\mathrm{N}$, and $^{15}\mathrm{O}$ RIBs via the $^{1}\mathrm{H}(^{7}\mathrm{Li}, ^{7}\mathrm{Be})n$, $^{2}\mathrm{H}(^{15}\mathrm{N}, ^{16}\mathrm{N})p$, and $^{1}\mathrm{H}(^{15}\mathrm{N}, ^{15}\mathrm{O})n$ inverse kinematics reactions, yielding purities of 85\%, 99\%, and 95\%, with intensities of $1.02\times10^{6}$, $2.7\times10^{5}$, and $1.0\times10^{5}$ pps, respectively. A $^{93m}\mathrm{Mo}$ isomer beam was also produced via the $\mathrm{^4He(^{94}Zr,} 5n)^{93m}\mathrm{Mo}$ reaction, achieving an intensity of $5.38\times10^{3}$ pps and a purity of 20\% (which can be further improved to $\sim$50\% with offline time-of-flight gating). By delivering a broader range of high-intensity secondary RIBs, this setup establishes a robust platform at RIBLL for low- and medium-energy nuclear astrophysics and reaction studies.
\end{abstract}


\end{frontmatter}


\section{Introduction}

High-intensity, high-purity radioactive ion beams (RIBs) are essential tools for modern nuclear physics, driving frontier research in exotic nuclear structure, isospin dynamics, and nuclear astrophysics \cite{geissel2005radioactive, brown2008nuclear, tsang2011isospin, iliadis2007nuclear}. In particular, measuring the extremely small cross sections of thermonuclear reactions at astrophysical Gamow window energies demands intense RIBs at very low energies \cite{iliadis2007nuclear, schatz2006xray, glorius2023lowenergy}. Historically, mainstream RIB facilities have predominantly relied on the Projectile-Fragmentation (PF) mechanism \cite{geissel2005radioactive, nilsson1999exotic}. However, applying PF to low-energy experiments (typically 1--5 MeV/nucleon) requires thick mechanical degraders to forcibly decelerate the energetic secondary beams. This deceleration process induces severe energy straggling, large angular divergence, and a severe attenuation of beam intensity \cite{geissel2005radioactive, lennard2002energy}, making it highly inefficient for low-energy precision measurements.

To overcome this physical bottleneck, experimentalists employ in-flight transfer reactions or inverse-kinematics fusion-evaporation reaction mechanisms through direct bombardment of gas targets with low-energy primary beams. Liquid-nitrogen-cooled ($\mathrm{LN_2}$-cooled) gas targets were introduced early on by Tribble \textit{et al.} \cite{tribble1989_MARS} for the production of radioactive secondary beams. Utilizing the Momentum Achromat Recoil Spectrometer (MARS) at the Texas A\&M University Cyclotron Institute, high-intensity and low-energy beams such as $^{11}\mathrm{C}$, $^{12,13}\mathrm{N}$, $^{13,14,15}\mathrm{O}$, $^{17}\mathrm{F}$, $^{20}\mathrm{Mg}$, and $^{20}\mathrm{Na}$ \cite{Shidling2018_21Na} have been successfully produced. In this configuration, the target gas ($\mathrm{H_2}$ or $\mathrm{^{3,4}He}$) is cooled down to liquid nitrogen ($\mathrm{LN}_2$) temperature, with the operating pressure typically maintained in the range of about 300 Torr to 1 atm. Yamaguchi \textit{et al.} \cite{yamaguchi2008} operated a $\mathrm{LN_2}$-cooled target at 85--90 K and produced a $^7$Be secondary beam of $2 \times 10^8$ pps with 74.6\% purity using a 5.6 MeV/u, 2.7 $\mu$A $^7$Li$^{3+}$ primary beam at the Center for Nuclear Study (CNS) Radioactive Ion Beam separator (CRIB). In that work, forced gas circulation was used to suppress beam-induced thermal rarefaction, reducing density loss from 30\% to 5\% for a heat deposition of 65 mW/mm at a flow rate of 55 L/min, demonstrating the effectiveness of cryogenic gas-target operation.

In order to obtain low-energy and high-intensity RIBs at the Radioactive Ion Beam Line in Lanzhou (RIBLL) \cite{ribll1_1999}, which is operated at the Heavy Ion Research Facility in Lanzhou (HIRFL) \cite{xia2002heavy} of the Institute of Modern Physics (IMP), Chinese Academy of Sciences, He \textit{et al.} \cite{he2012} previously demonstrated an alcohol-cooled gas target, which produced a $^{22}$Na beam with $\sim$30\% purity and $1.7 \times 10^4$ pps intensity via the $^1\text{H}(^{22}\text{Ne}, ^{22}\text{Na})n$ reaction at operating  temperature $2^\circ\text{C}$. This represented an order-of-magnitude intensity gain over PF schemes. Although alcohol-cooled target \cite{he2012} provided an earlier gas-target implementation at RIBLL, advancing to cryogenic $\mathrm{LN_2}$-cooled operation, which would yield an approximately 3-fold increase in target thickness at the same pressure, was pursued to increase the intensity of RIB. We have developed and installed a $\mathrm{LN_2}$-cooled cryogenic gas target system at RIBLL, the design of which has benefited greatly from the pioneering work of Yamaguchi \textit{et al.} \cite{yamaguchi2008} and He \textit{et al.} \cite{he2012}.

This paper details the technical design of the cryogenic gas target system, along with the beam line and detector configurations. The beam purification and measurement processes are explicitly validated using newly acquired particle identification (PID) spectra. The successful commissioning of this system provides a robust platform for producing low- and medium-energy RIBs at RIBLL for nuclear astrophysics and reaction mechanism studies.

\section{Design and development of the cryogenic gas target}

\subsection{Overall structure and cryogenic cooling}

\begin{figure}[htbp]
	\centering
	\includegraphics[width=0.49\textwidth]{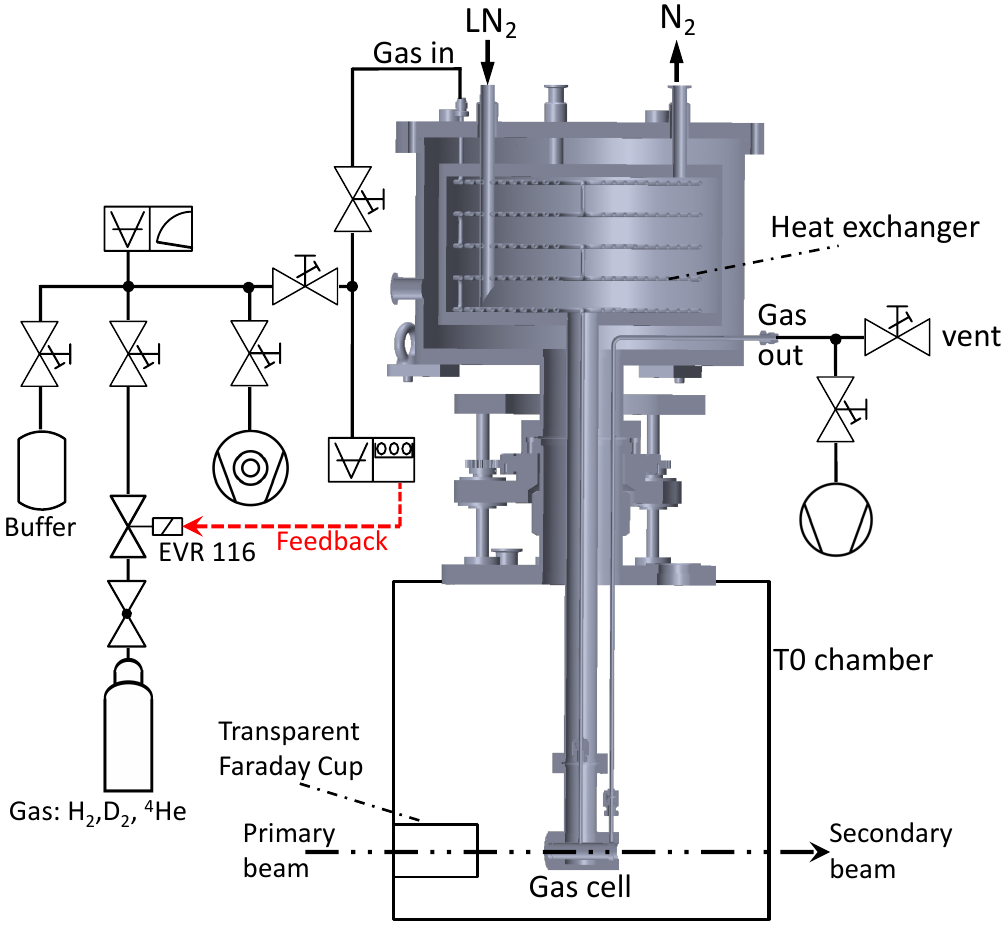}
	\caption{The diagram of the present cryogenic gas target system at HIRFL-RIBLL. The cross-sectional view of the cryogenic gas target is shown.}
	\label{fig_setup}
\end{figure}

The cryogenic gas target system is installed in the $\mathrm{T}_0$ target chamber of the RIBLL beam line, the core component of which is a cylindrical gas cell measuring $80\ \mathrm{mm}$ in length with an internal diameter of $20\ \mathrm{mm}$. The detailed design of the present gas target is shown in Fig. \ref{fig_setup}. To effectively enhance the gas target density, a dedicated Dewar vessel with a capacity of approximately $18.5\ \mathrm{L}$ is equipped at the top of the system. The $\mathrm{LN}_2$ level decreases substantially over approximately 12 hours; in routine operation, the Dewar was refilled every 8 hours. A $\mathrm{LN}_2$-cooled loop is utilized to deeply cool the light-mass reaction gas filled in the target cell to cryogenic temperature. While a thermocouple attached to the gas-cell outlet typically recorded temperatures of 82 to 86 K during beam irradiation, localized beam energy deposition possibly increases the actual gas temperature within the beam-interaction zone. The monitored outlet temperatures are slightly higher than the boiling point of $\mathrm{LN}_2$ (77 K) due to unavoidable thermal conduction and radiation heat exchange between the gas cell and the ambient environment, as well as continuous heat deposition from the primary beam passing through the target.

To maximize heat exchange efficiency and ensure uniform gas cooling, the gas-carrying copper tubing is configured in a five-layer helical coil structure immersed in the $\mathrm{LN}_2$ bath, with ten turns per layer (Fig.~\ref{fig_setup}). Under the same operating pressure (up to $\sim1500\ \mathrm{mbar}$ for present design), this cryogenic condition increases the gas areal density to approximately 3.0 times that achieved by alcohol-cooled device \cite{he2012}, which was operated at $2^\circ\text{C}$ instead of its designed minimum operating temperature of $-30^\circ\text{C}$. 

\subsection{Gas cell and cryogenic vacuum sealing technique}

\begin{figure}[htbp]
	\centering
	\includegraphics[width=0.49\textwidth]{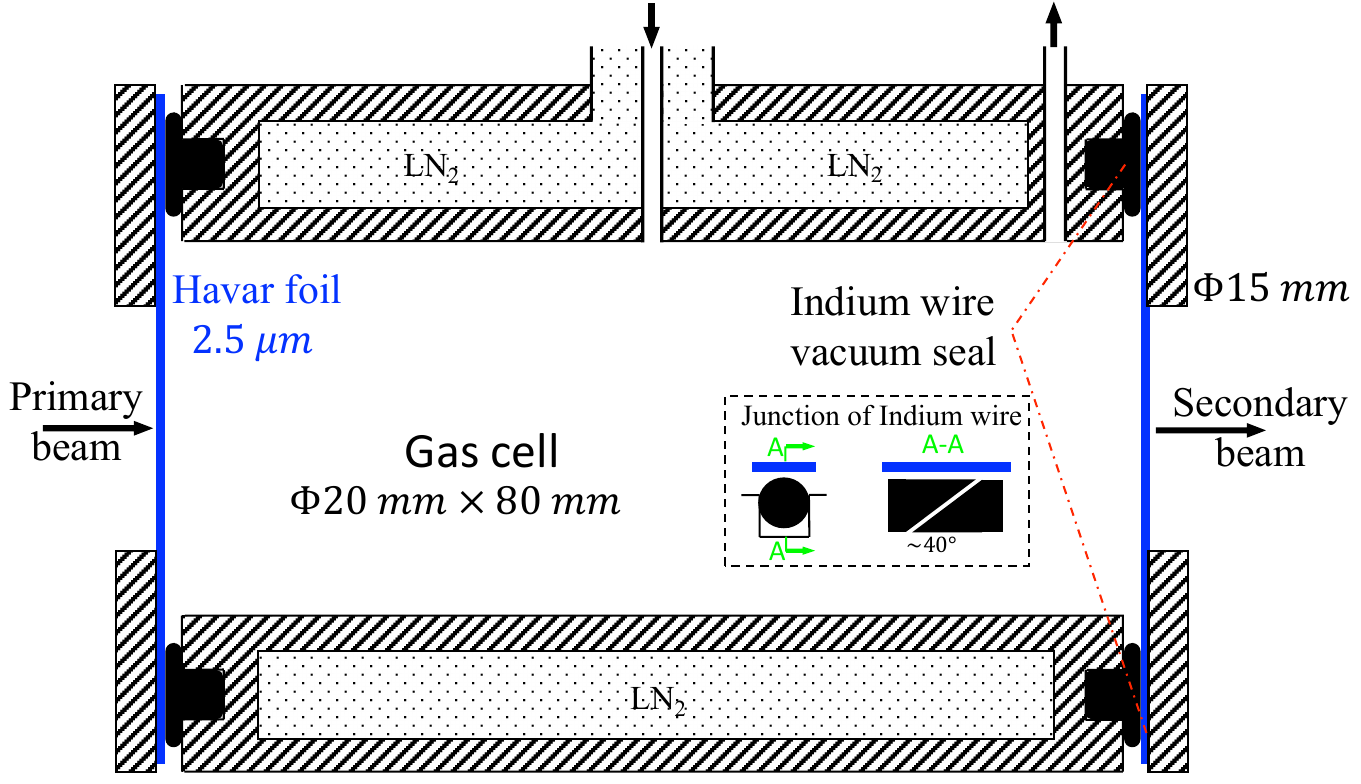}
	\caption{Cross-sectional view of the gas cell for the present gas target with key components labeled. The dimensions shown in the figure are not to scale.}
	\label{fig_gas_cell}
\end{figure}

The structure of the critical gas cell is displayed in Fig. \ref{fig_gas_cell}. 
The gas target cell is machined from stainless steel to ensure high mechanical rigidity and dimensional stability at cryogenic temperatures. This choice was primarily driven by the requirements of the indium-wire vacuum seal and the ultrathin Havar-window clamping structure. The indium seal requires a rigid annular groove to maintain uniform compression during assembly and cryogenic cycling, while the 2.5-$\mu$m Havar foil requires a flat and mechanically stable clamping surface to avoid local stress concentration. Although copper has a much higher thermal conductivity than stainless steel, a simple thermal-resistance estimate shows that the temperature drop across the 2-mm stainless-steel wall is negligible under the present heat load. For the maximum calculated gas heat deposition of 1.29 W, the expected temperature drop across the stainless-steel wall is of the order of 0.1 K, much smaller than the possible temperature non-uniformity in the gas itself. Therefore, the stainless-steel cell provides the required mechanical robustness without becoming the dominant thermal bottleneck.

Considering the harsh operating environment characterized by cryogenic temperatures and high pressures, high-strength Havar foils with a thickness of only $2.5\ \mu\mathrm{m}$ are employed as the beam entrance and exit windows at both ends of the gas cell. This provides excellent mechanical strength to withstand the pressure difference between the target gas and the beam line vacuum while minimizing the energy straggling of the beam.

\begin{figure}[htbp]
	\centering
	\includegraphics[width=0.49\textwidth]{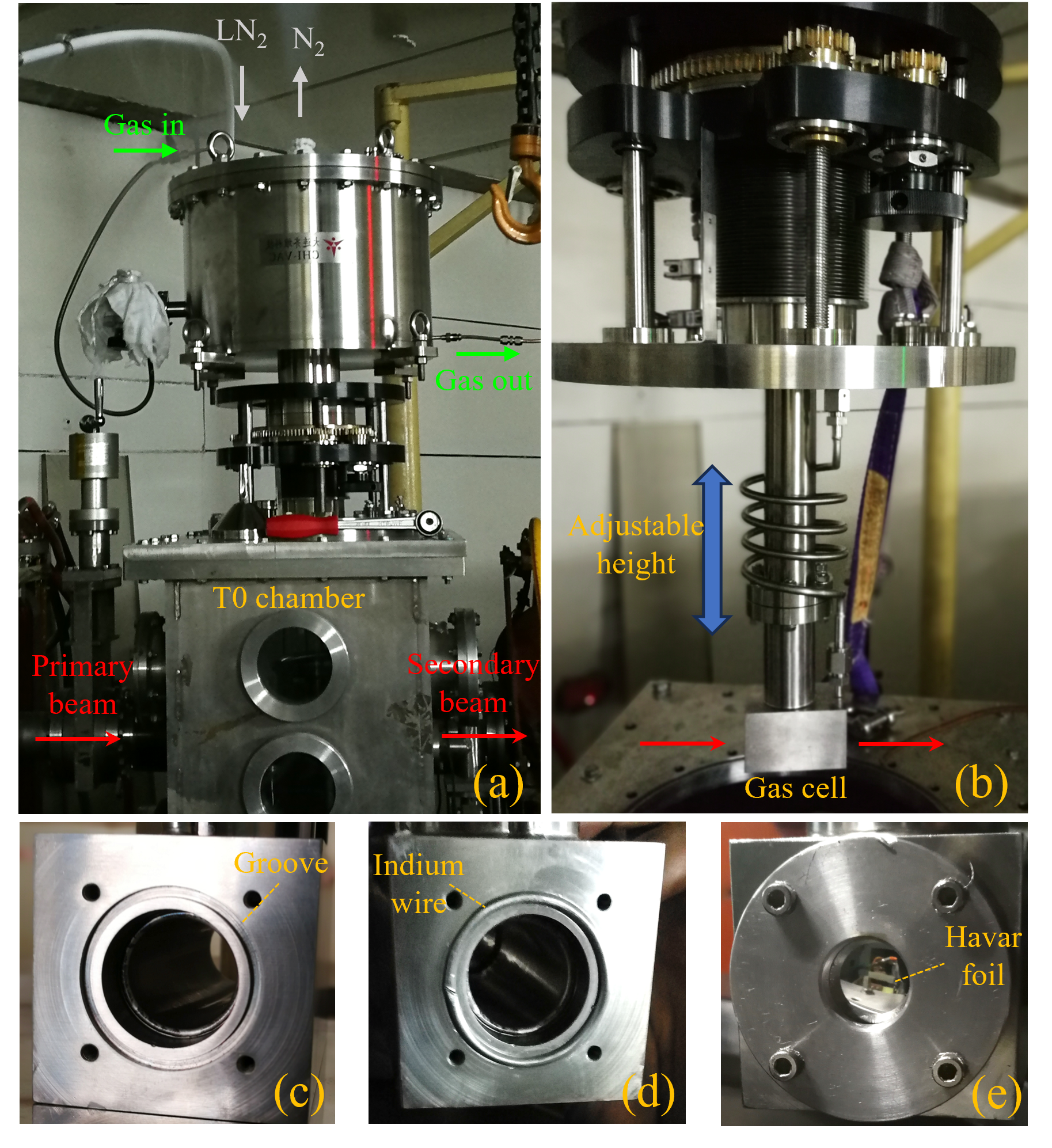}
	\caption{Photographs of the present cryogenic gas target system at RIBLL. (a) The overall target assembly mounted on the T0 vacuum chamber, indicating the gas and cryogenic fluid feedthroughs, as well as the beam axis. (b) The inner assembly featuring the adjustable height mechanism and the gas cell. (c) The machined steel target cell flange showing the 1.9-mm-wide annular sealing groove. (d) A 2-mm-diameter pure indium wire placed within the groove for cryogenic vacuum sealing. (e) The fully assembled target window with the $2.5\ \mu\mathrm{m}$ Havar foil uniformly compressed by the clamping ring. }
	\label{fig_gas_target_setup}
\end{figure}

To achieve reliable leak-tight operation under the combined conditions of cryogenic temperature, internal pressure up to 1500 mbar, and prolonged beam irradiation, a specialized cryogenic vacuum sealing technique was implemented (Fig. \ref{fig_gas_target_setup}). Traditional rubber O-rings, which are prone to hardening and failure at low temperatures, were replaced by high-purity indium wire. As shown in Fig. \ref{fig_gas_target_setup}(a) and (b), the gas cell is integrated into the vertical cryostat assembly. The target cell flange features a precision annular sealing groove (1.9 mm wide, 1.5 mm deep) to securely house the 2-mm-diameter indium wire (Fig. \ref{fig_gas_target_setup}(c) and (d)). To eliminate potential leakage paths, the two ends of the indium wire were cut at a $40^{\circ}$ bevel angle to form a seamless closed ring. A critical mechanical challenge during assembly is preserving the structural integrity of the ultrathin Havar foil (Fig. \ref{fig_gas_target_setup}(e)). The foil is temporarily affixed to the surface of an Aluminum clamping ring (featuring a $\Phi$15 mm smoothed round hole in the center) using a thin layer of vacuum grease to help maintain a flat, wrinkle-free surface during assembly. This clamping ring is then tightened using four screws via a strictly controlled, uniform diagonal compression procedure. This technique gradually deforms the indium wire flush with the groove surface, thereby minimizing localized stress concentrations that could rupture the fragile window. The Havar foils are thermally anchored through the uniformly compressed clamping ring and the steel flange, providing a conductive path from the beam-heated window region to the LN$_2$-cooled target body.

\subsection{Dynamic gas handling and pressure regulation}

The target volume inside the cell is connected to the gas-handling system via dedicated inlet and outlet channels, enabling controlled gas renewal and pressure regulation. A Pfeiffer Vacuum EVR 116 electromagnetic gas control valve is employed together with an RVC 300 control unit for pressure regulation over 100--1500 mbar. The pressure control accuracy is approximately $\pm$2 mbar, which limits the pressure contribution to the uncertainty of the nominal areal density.

Considering space and budget constraints, a simplified dynamic micro-flow constant-pressure strategy was implemented to enable stable extended operation without the added complexity of a forced circulation system. Since the electromagnetic gas control valve provides gas injection only, a small continuous exhaust is required to enable active feedback regulation against slow pressure drifts caused by thermal changes in the gas-handling lines. In operation, the exhaust valve was slightly opened to maintain a small steady flow rate of approximately 100--150 mL/min at room temperature. This micro-flow was used for pressure feedback regulation and gas renewal, and it also helped purge possible impurities. It was not used as a forced-circulation cooling mechanism. Accordingly, the target thicknesses quoted in this work are treated as nominal target thicknesses derived from the monitored pressure and nominal cryogenic operating temperature, rather than as direct measurements of the local gas-density distribution in the beam-interaction region.

\subsection{System reliability and in-beam validation}

To verify the overall gas tightness and operational reliability of the gas target system, rigorous static pressure-holding tests were conducted. Initially, under room-temperature conditions, the gas target was pressurized with reaction gas to 1000~mbar and 1500~mbar, respectively. Once pressurized, all gas inlet and outlet valves of the target cell were completely sealed, while the external T0 target chamber was maintained under a high-vacuum environment. Over a continuous monitoring period of $\sim$8~hours, the system pressure remained highly stable, and no obvious pressure drop or gas leakage was observed. Subsequently, the sealing performance was further evaluated under cryogenic operating conditions with $\mathrm{LN_2}$. After pressurizing and fully sealing the target cell, no significant gas leakage was detected over an hour-long period.

Following the static pressure tests, the system was subjected to a more stringent dynamic validation during actual in-beam operations. During beam bombardment, the actual maximum operating pressure of the gas target reached 1000 mbar. Even under the high thermal load imposed by a $^{7}\mathrm{Li}^{3+}$ primary beam with an energy of 8.8~MeV/u and a beam current as high as 1.6~$\mathrm{\mu A}$, the system maintained excellent gas tightness, and no gas leakage was detected. Furthermore, the Havar foils maintained mechanical integrity and leak-tight performance throughout day-long high-intensity irradiation campaigns, with no rupture or structural failure observed. This demonstrates the gas tightness, mechanical robustness, and operational reliability of the cryogenic gas target under the tested in-beam conditions.

\section{Experimental Results of RIBs Production}

\subsection{The setup of beam line}

\begin{figure*}
	\centering
	\includegraphics[width=0.92\textwidth]{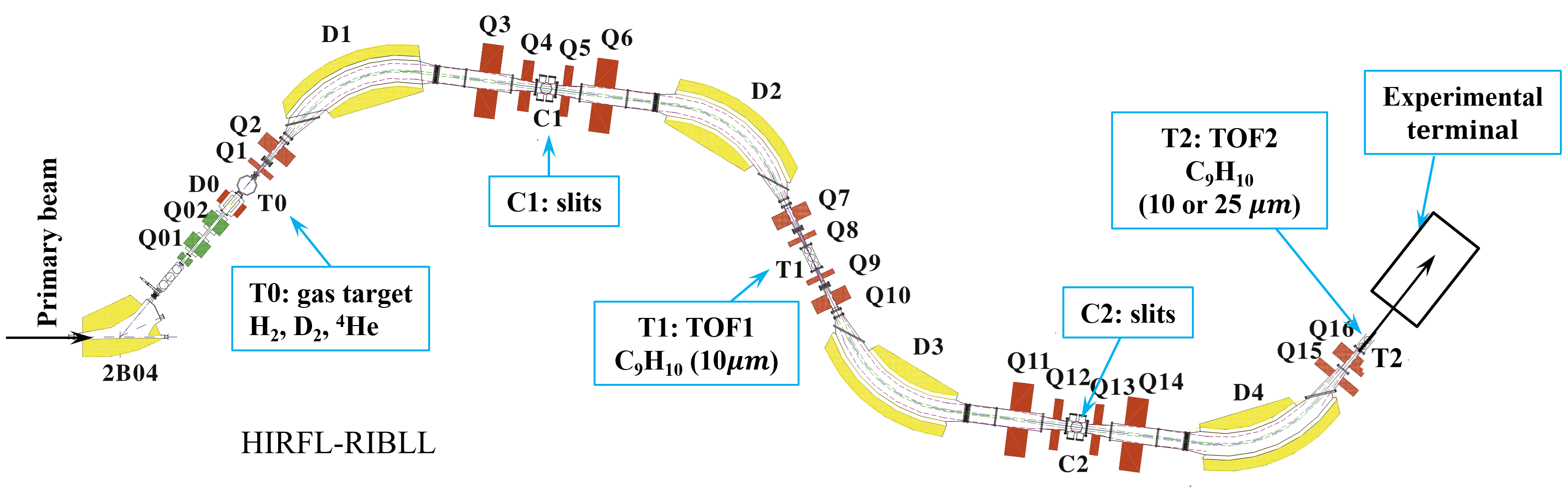}
	\caption{Schematic of HIRFL-RIBLL  beam line and the experimental setup for production of the secondary $^7$Be, $^{15}$O, $^{16}$N, and $^{93m}$Mo beams by employing present $\mathrm{LN}_2$- cooled cryogenic gas target.}
	\label{fig_schematic}
\end{figure*}

All experiments in this work were conducted at the HIRFL-RIBLL beam line of IMP. The layout of the beam line is illustrated in Fig.~\ref{fig_schematic}. The primary heavy-ion beam accelerated by the HIRFL cyclotron (SFC or SSC) is transported to the T0 target chamber, where it bombards the $\mathrm{LN}_2$-cooled cryogenic gas target (H$_2$, D$_2$, or $^4$He) under inverse kinematics to produce radioactive secondary nuclei.

The 35-meter-long RIBLL spectrometer features a double-achromatic anti-symmetric optical design with a maximum magnetic rigidity of 2.8 T$\cdot$m, which covers all secondary beams produced in this work. The reaction products are separated by magnetic rigidity through four dipole magnets (D1, D2, D3, D4), with two momentum-defining slits (C1 and C2) selecting the desired rigidities and providing further purification. Finally, purified secondary beam is delivered to the experimental terminal.

Detectors based on time-of-flight (TOF) and energy loss ($E$ or $\Delta E$-$E$) are configured at the beam line for precise particle identification. Ultra-thin plastic scintillators ($\mathrm{C_9H_{10}}$) are installed at focal points T1 and T2 as time pick-up detectors. TOF1 has a thickness of 10~$\mu\mathrm{m}$, whereas TOF2 employs either a 10-$\mu\mathrm{m}$ foil for the $^{7}\mathrm{Be}$ and $^{93m}\mathrm{Mo}$ measurements, or a 25-$\mu\mathrm{m}$ foil for the $^{16}\mathrm{N}$ and $^{15}\mathrm{O}$ beams. Utilizing the exceptionally long flight path of approximately 16.8~m between these two points, the TOF system achieves nanosecond-level time resolution for high-energy ions. Furthermore, during initial beam tuning or in low-count-rate modes, a large-area thick silicon detector (300~$\mu\mathrm{m}$ in thickness, energy resolution better than 2\%) is positioned behind the T2 focal point to measure the total kinetic energy ($E$) of the particles. By combining the TOF signals with the energy signals from the Si detector to construct a two-dimensional phase-space projection (TOF-$E$ plot), unambiguous PID is achieved. Upon entering the formal high-intensity measurement phase, the silicon detector is removed from the beam line to prevent severe radiation damage, leaving only the TOF system for online monitoring.

In formal experiments, two parallel-plate avalanche counters (PPACs) \cite{ma2011ppac} are frequently installed downstream of TOF2 to provide beam position and angle information via extrapolation of their two-dimensional hit data. These detectors have a 1 mm position resolution and $\sim$0.3 ns time resolution.

\subsection{$^{16}\mathrm{N}$ and $^{15}\mathrm{O}$ Commissioning Beams}

\begin{table*}
	\centering
	\caption{Radioactive ion beams produced using the present $\mathrm{LN}_2$-cooled cryogenic gas target at RIBLL.}
	\label{Tab_summary_present_RIB}
	\setlength{\tabcolsep}{3pt} 
	\begin{threeparttable}  
	\begin{tabular}{ccccccccccc} 
		\toprule
		\multirow{2}{*}{No.} & \multicolumn{4}{c}{Secondary beam} & \multicolumn{3}{c}{Gas target} & \multicolumn{3}{c}{Primary beam} \\ 
		\cmidrule(lr){2-5} \cmidrule(lr){6-8} \cmidrule(lr){9-11} 
		& Ion & E (MeV/u) & Int. (pps) & Purity & Gas & P (mbar) & Thk.\tnote{\dag} (mg/cm$^2$) & Ion & E (MeV/u) & Current (nA) \\
		\midrule
		1 & $^{16}$N$^{7+}$ & $\sim$6.4 & $2.7 \times 10^5$ & >99\% & D$_2$ & 530 & $\sim$2.45 & $^{15}$N$^{7+}$ & 8.5 & 300 \\
		2 & $^{15}$O$^{8+}$ & $\sim$7.5  & $1.0 \times 10^5$ & 95\% & H$_2$ & 800 & $\sim$1.85 & $^{15}$N$^{7+}$ & 9.5 & 550 \\		
		3 & $^{7}$Be$^{4+}$ & 7.21 & $1.02 \times 10^6$ & 85\%\tnote{\ddag} & H$_2$ & 1000 & $\sim$2.31 & $^{7}$Li$^{3+}$ & 8.8 & 1600 \\
		4 & $^{7}$Be$^{4+}$ & 6.86 & $3.24 \times 10^5$ & 90\% & H$_2$ & 800 & $\sim$1.85  & $^{7}$Li$^{3+}$ & 8.6 & 870 \\
		5 & $^{93m}$Mo$^{37,38+}$ & 11.6 & $5.38 \times 10^3$ & 20\%\tnote{\S} & $^4$He & 980 & $\sim$4.5 & $^{94}$Zr$^{19+}$ & 16.7 & 100 \\  
		\bottomrule
	\end{tabular}
	\begin{tablenotes}    
		\footnotesize               
		
		\item[\dag] The values represent nominal target thicknesses calculated from the ideal-gas law using the monitored pressure and a nominal cryogenic operating temperature of 84 K. No correction for the local beam-induced temperature distribution is applied.
		\item[\ddag] A better purity of $\sim$90\% with a lower $^{7}$Be$^{4+}$ intensity of 7.0 $\times$ $10^5$ pps.
		\item[\S] An effective purity of $\sim$50\% could be achieved by applying offline TOF gating.    
	\end{tablenotes}   
	\end{threeparttable}  
\end{table*}

The $^{16}\mathrm{N}$ and $^{15}\mathrm{O}$ secondary beams were firstly produced RIBs to commission present cryogenic gas target system. Both secondary beams were produced utilizing the exact same type of primary beam, $^{15}\mathrm{N}^{7+}$, but at different incident energies and with different target gases, as summarized in Table~\ref{Tab_summary_present_RIB}.

\begin{figure}
	\centering
	\includegraphics[width=0.5\textwidth]{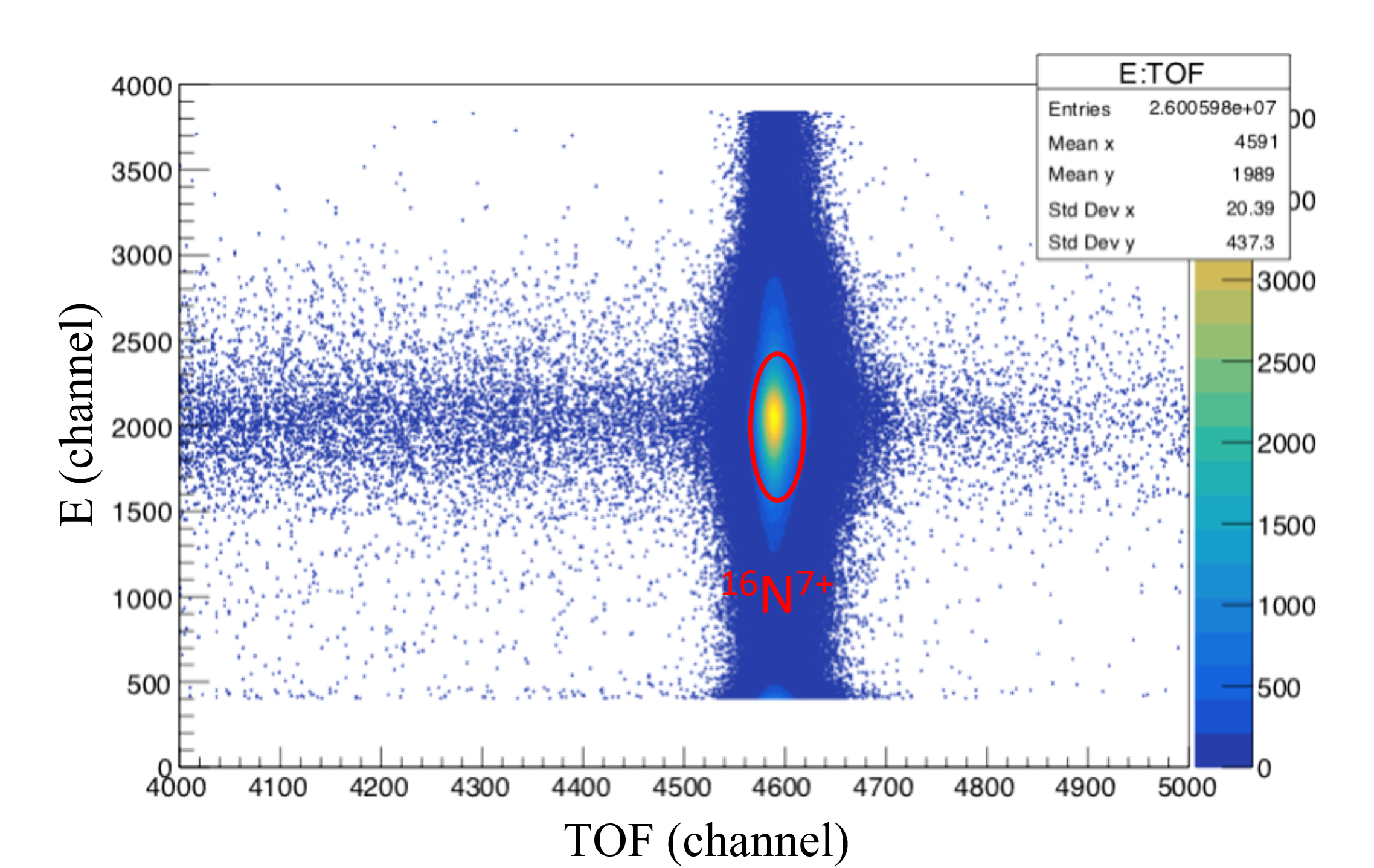} 
	\caption{E vs TOF particle identification spectrum for the $^{16}$N secondary beam produced by $^{2}\mathrm{H}(^{15}\mathrm{N},p)^{16}\mathrm{N}$ reaction using D$_2$ target at 530 mbar.}
	\label{fig:E_TOF_16N}
\end{figure}

For production of the $^{16}\mathrm{N}$ beam, an 8.5 MeV/u $^{15}\mathrm{N}^{7+}$ primary beam bombarded a cryogenic $\mathrm{D_2}$ gas target maintained at approximately 530~mbar. The C1 and C2 slits were set to be $\pm 5$ mm and $\pm 20$ mm, respectively. Through the $^{2}\mathrm{H}(^{15}\mathrm{N}, ^{16}\mathrm{N})p$ transfer reaction, a $\sim$6.4 MeV/u $^{16}\mathrm{N}^{7+}$ secondary beam was obtained. As shown in the TOF-$E$ PID spectrum (Fig.~\ref{fig:E_TOF_16N}), the target isotope was clearly resolved, achieving a $^{16}\mathrm{N}$ beam with in-flight purity of $>99\%$ and an intensity of $\sim2.7 \times 10^5$~pps with a 300 nA primary beam current. During initial commissioning, the beam purity was limited by unoptimized magnetic rigidity settings and collimation conditions. After systematic optimization, the final in-flight purity of $^{16}\mathrm{N}$ exceeded 99\%.

The quality of the secondary RIBs was systematically characterized. During initial commissioning runs, magnetic rigidity scanning demonstrated that the secondary beams exhibit a well-defined Gaussian distribution with a typical momentum spread $\Delta p/p$ of approximately 0.4\%.

Under these relatively low beam-current conditions, beam-induced density reduction is expected to be small. We varied the current of the $^{15}$N$^{7+}$ primary beam over the range of 120--300 nA and simultaneously measured the secondary beam counts at T1 and T2 focal planes. The transmission efficiency from T1 to T2 was approximately 20\%, but it fluctuated from 10\% to 27\%. This value is typical for RIBLL operations and represents a deliberate trade-off between beam intensity and purity. The transmission loss mainly arises from the magnetic rigidity selection by the D3 and D4 dipole magnets and the collimation by the C2 slits.

\begin{figure}
	\centering
	\includegraphics[width=0.5\textwidth]{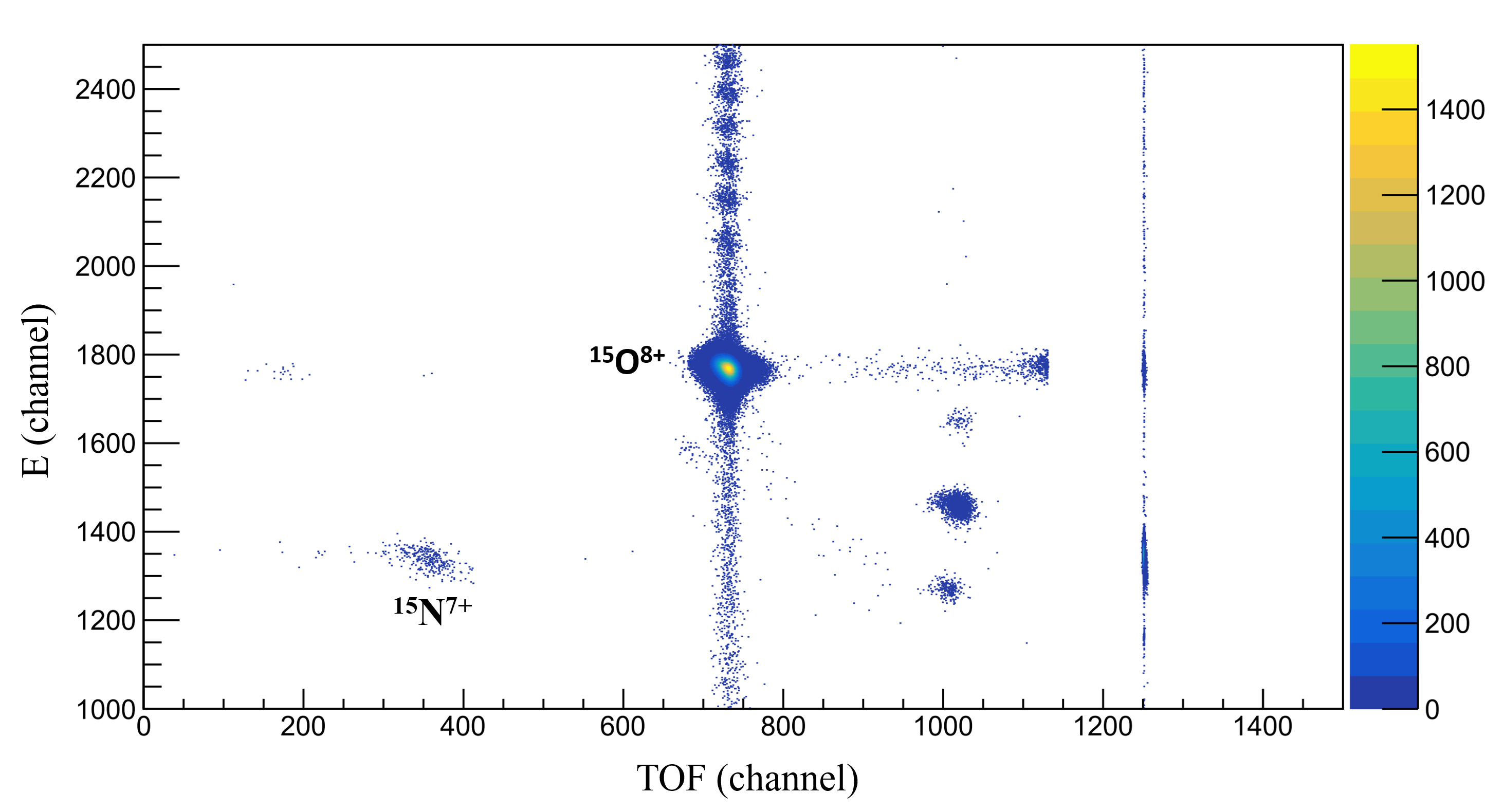} 
	\caption{E vs TOF particle identification spectrum for the $^{15}$O secondary beam produced by  $^{1}\mathrm{H}(^{15}\mathrm{N}, n)^{15}\mathrm{O}$ reaction with H$_2$ target at 800 mbar.}
	\label{fig:E_TOF_15O}
\end{figure}

Similarly, the $^{15}\mathrm{O}$ beam was produced via the  $^{1}\mathrm{H}(^{15}\mathrm{N}, ^{15}\mathrm{O})n$ reaction. In this case, a higher primary beam energy of 9.5~MeV/u and a cryogenic $\mathrm{H_2}$ gas target at 800 mbar were employed. The C1 and C2 slits were set to be $\pm 2$ mm and $\pm 20$ mm, respectively. The resulting $\sim$7.5 MeV/u $^{15}\mathrm{O}^{8+}$ beam reached an intensity of $\sim1.0 \times 10^5$ pps with a 550 nA primary beam, and a purity of 95\% was confirmed by the PID spectrum (Fig.~\ref{fig:E_TOF_15O}).

These specific low-energy RIBs had not been routinely available at RIBLL through the PF method, due to the severe beam attenuation during the thick-degrader deceleration process. The secondary beam intensities reported here were obtained during the commissioning phase. Higher intensities can be readily achieved by further increasing the primary beam current and optimizing the target gas pressure.

\subsection{$^7\mathrm{Be}$ Beam}

\begin{figure}
	\centering
	\includegraphics[width=0.5\textwidth]{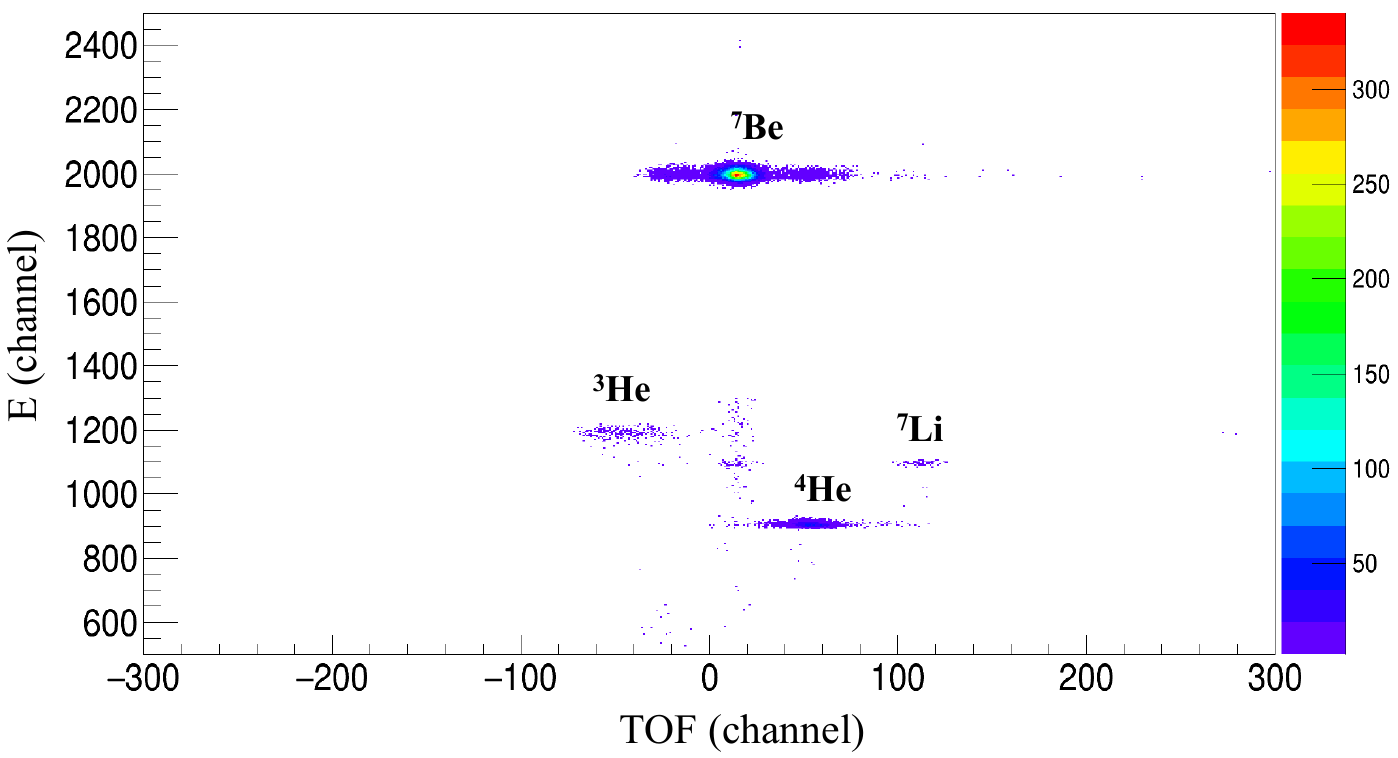}
	\caption{E vs TOF particle identification spectrum for the $^{7}$Be secondary beam produced by $^{1}\mathrm{H}(^{7}\mathrm{Li}, n)^{7}\mathrm{Be}$ reaction using H$_2$ target at 1000 mbar.}
	\label{fig_7Be_E_TOF}
\end{figure}

\begin{figure}[htbp]
	\centering
	\includegraphics[width=0.5\textwidth]{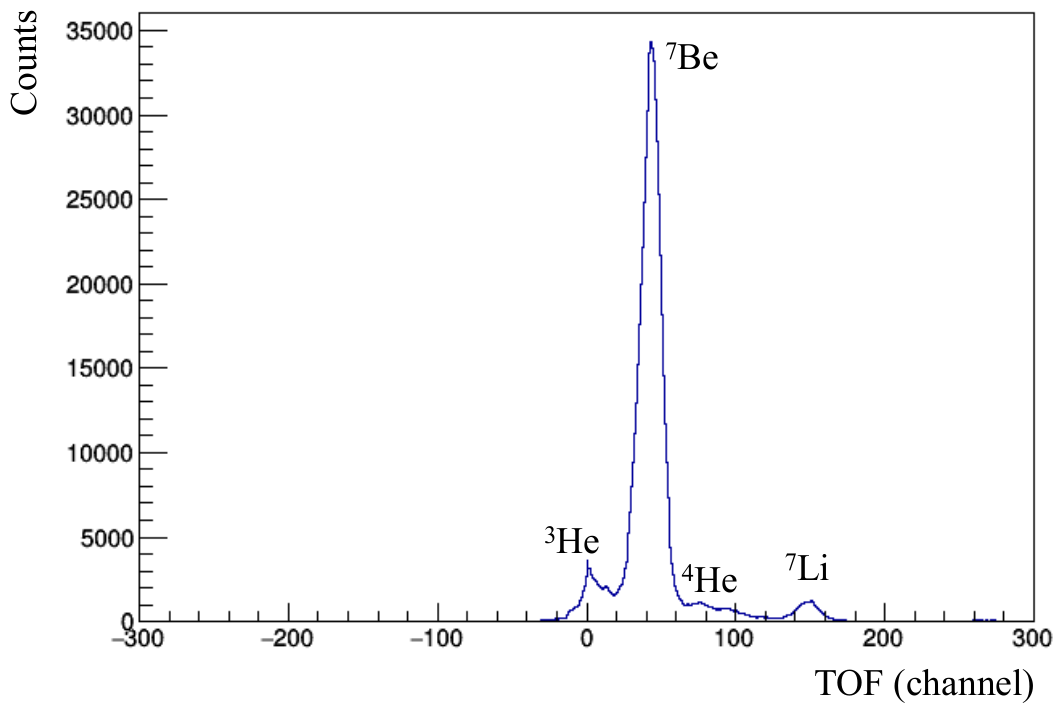}
	\caption{TOF spectra of $^{7}$Be secondary beam production in Fig. \ref{fig_7Be_E_TOF}.}
	\label{fig_TOF_7Be}
\end{figure}

The $^{7}\mathrm{Be}$ secondary beam was generated via the $^{1}\mathrm{H}(^{7}\mathrm{Li}, ^{7}\mathrm{Be})n$ reaction using the cryogenic $\mathrm{H_2}$ gas target. To demonstrate the maximum production capability of the system, a $^{7}\mathrm{Li}^{3+}$ primary beam with an energy of 8.8~MeV/u and a high intensity of 1.6~$\mathrm{\mu A}$, delivered by the HIRFL-SFC, bombarded the gas target. The C1 and C2 slits were set to be $\pm 10$ mm and $\pm 15$ mm, respectively. Following separation and purification by the RIBLL spectrometer, the $^{7}\mathrm{Be}$ beam achieved a purity of 85\% and an intensity of $\sim1.02 \times 10^6$ pps (a better purity of $\sim$90\% with a lower intensity of $7.0 \times 10^5$ pps), summarized in Table \ref{Tab_summary_present_RIB}. 
The TOF-$E$ PID spectrum for the $^{7}\mathrm{Be}$ beam under these conditions is displayed in Fig. \ref{fig_7Be_E_TOF}, and TOF spectrum is shown in Fig. \ref{fig_TOF_7Be}. This achievement marks the first time that the HIRFL-RIBLL facility has provided such a high-quality, high-purity $^{7}\mathrm{Be}$ secondary beam. A summary of $^7$Be beams produced by different facilities is listed in Table \ref{Tab_summary_7Be}.

\begin{table*}
	\centering
	\caption{Summary of $^7$Be beam production.}
	\label{Tab_summary_7Be}
	\setlength{\tabcolsep}{6pt}
	\begin{tabular}{p{0.1cm} p{1.9cm} p{0.8cm} p{1.5cm} p{0.8cm} p{0.8cm} p{1.8cm} p{1.7cm} p{1.2cm} p{0.6cm} p{0.8cm}}
		\toprule
		& Facility  & Primary beam & Accelerator  & E$_{beam}$ (MeV/u) & I$_{beam}$ ($\mu$A) & Primary target & Thickness (mg/cm$^2$) & $^7$Be (pps) & Purity & Ref.                                     \\  
		\midrule
		1  & LNL & $^7$Li & tandem & 4.9 & 0.26 & H$_2$ &  & 2.5$\times$10$^5$ & 99\%  & \cite{mazzocco2015} \\ \hline
		2  & CNS & $^7$Li & cyclotron & 5.6 & 2.7 & H$_2$ & 2.3 & 2.0$\times$10$^8$ & 74.6\% & \cite{yamaguchi2008} \\ \hline
		3  & CNS & $^7$Li & cyclotron & 8.6 & 1.0 & H$_2$ & 2.3 & 5.0$\times$10$^5$ & 90\% & \cite{yamaguchi2008} \\ \hline
		4  & Leuven  & $^7$Be  & cyclotron  &  &  &  &  & 5.0$\times$10$^6$ & 100\%  & \cite{raabe2006} \\  \hline
		5  & Notre Dame & $^6$Li  & tandem  & 5.5 &   & $^3$He   &  & 7.3$\times$10$^4$ &  & \cite{aguilera2009}  \\  \hline
		6  & Notre Dame & $^6$Li  & tandem  & 6.17 & 0.5 & $^3$He &  & 1.0$\times$10$^5$ &  & \cite{amro2007} \\ \hline
		7  & RIBRAS (Brazil) & $^6$Li &   & 4.7 & 1 & $^3$He & 1 atm & 4.0$\times$10$^5$ &   & \cite{zamora2011}  \\  \hline
		8  & HIRA (IUAC India) & $^7$Li  & tandem & 3.3 &  & polypropylene (CH$_2$)n & 12 $\mu$m & 1.0$\times$10$^4$ & 99\% & \cite{kalita2006} \\ \hline
		\multirow{2}{*}{9} & \multirow{2}{*}{IMP RIBLL} & \multirow{2}{*}{$^7$Li} & \multirow{2}{*}{cyclotron} & \multirow{2}{*}{8.8} & \multirow{2}{*}{1.6} & \multirow{2}{*}{H$_2$} & \multirow{2}{=}{$\sim$2.31 \\ (1000 mbar)} & 1.02$\times$10$^6$ & 85\% & \multirow{2}{=}{Present work} \\
		\cline{9-10}
		& & & & & & & & 7.0$\times$10$^5$ & 90\% & \\ 
		\bottomrule
	\end{tabular}
\end{table*}

However, subsequent experimental campaigns did not always require such high beam intensities. For instance, in a recent study of the $^{7}\mathrm{Be} + ^{120}\mathrm{Sn}$ reaction \cite{chang2023quasielastic,chang2025current}, the primary $^{7}\mathrm{Li}^{3+}$ beam energy was slightly decreased to 8.6~MeV/u, and the primary beam current was regulated within a lower range of 450--870 nA. Correspondingly, the energy of the extracted $^{7}\mathrm{Be}$ secondary beam was determined to be 48.05 MeV (6.86 MeV/u ). Under these tuning conditions, the secondary beam intensity was stably maintained between $1.62 \times 10^5$ and $3.24 \times 10^5$~pps with a purity of $\sim$90\%.

\begin{figure}[htbp]
	\centering
	\includegraphics[width=0.5\textwidth]{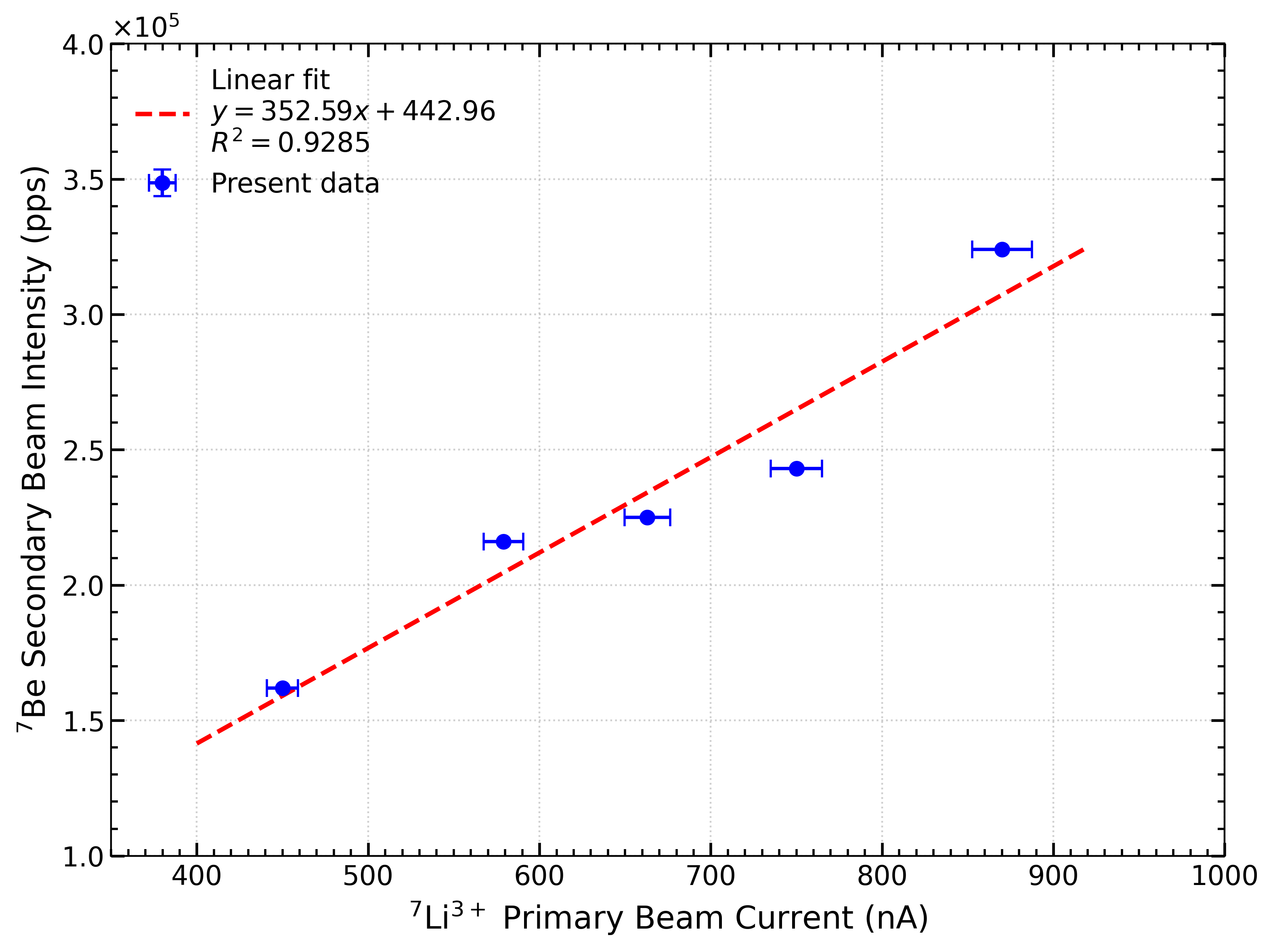}
	\caption{Absolute intensity of the $^{7}\mathrm{Be}$ secondary beam as a function of the $^{7}\mathrm{Li}^{3+}$ primary beam current. The blue circles represent the present experimental data points, and the red dashed line indicates the linear least-squares fit ($R^2 = 0.9285$).}
	\label{fig_7Be_intensity_linear_fit}
\end{figure}

The relationship between the $^7\mathrm{Be}$ secondary beam intensity and the $^7\mathrm{Li}^{3+}$ primary beam current is shown in Fig. \ref{fig_7Be_intensity_linear_fit}. A linear least-squares fit to the experimental data yields a slope of $352.6 \pm 20$ pps/nA and a coefficient of determination $R^2 = 0.9285$. The approximately linear trend is consistent with the absence of a large beam-induced target-density reduction within the tested current range up to 870 nA, supporting the operational stability of the cryogenic gas target under these conditions.

To elucidate the reaction dynamics of weakly bound nuclei near the Coulomb barrier, $^{7}\mathrm{Be}$ was selected as a key candidate. As a radioactive nucleus situated near the proton drip line with a half-life of 53.2 days, $^{7}\mathrm{Be}$ exhibits a low separation energy of 1.586 MeV, making it highly susceptible to breakup into $^{3}\mathrm{He}$ and $^{4}\mathrm{He}$ clusters. Furthermore, $^{7}\mathrm{Be}$ attracts broad interest because it serves as the core of the well-known proton-halo nucleus $^{8}\mathrm{B}$. A comprehensive understanding of the $^{7}\mathrm{Be}$ reaction mechanism is crucial for unraveling the complex dynamics of $^{8}\mathrm{B}$, which possesses an extremely low proton separation energy ($S_p = 0.138$~MeV) and plays a pivotal role in nuclear astrophysics. Utilizing the $\sim$7~MeV/u $^{7}\mathrm{Be}$ secondary beam produced by the present cryogenic target, elastic scattering angular distribution measurements for both the $^{7}\mathrm{Be} + ^{209}\mathrm{Bi}$ and $^{7}\mathrm{Be} + ^{120}\mathrm{Sn}$ systems have been performed.

\subsection{High-Spin Isomer $^{93m}\mathrm{Mo}$ Beam}

A key technical breakthrough has been achieved at the RIBLL regarding the development and application of the high-spin isomer beam $^{93m}\mathrm{Mo}$ \cite{guo2022probing,ding2026isomer}. In the early stages of the research \cite{guo2022probing}, limited by the primary beam energy, the fusion-evaporation reaction channel $^{12}\mathrm{C}(^{86}\mathrm{Kr}, 5n)^{93m}\mathrm{Mo}$ was adopted, which yielded a secondary beam energy of 4.95 MeV/u. Due to this low kinetic energy, standard detectors could not be implemented for TOF measurements. Furthermore, severe contamination from the extensively scattered primary beam resulted in an isomer beam intensity of 408 pps and a purity of 0.6\% \cite{guo2022probing}.

\begin{figure}
	\centering
	\includegraphics[width=0.5\textwidth]{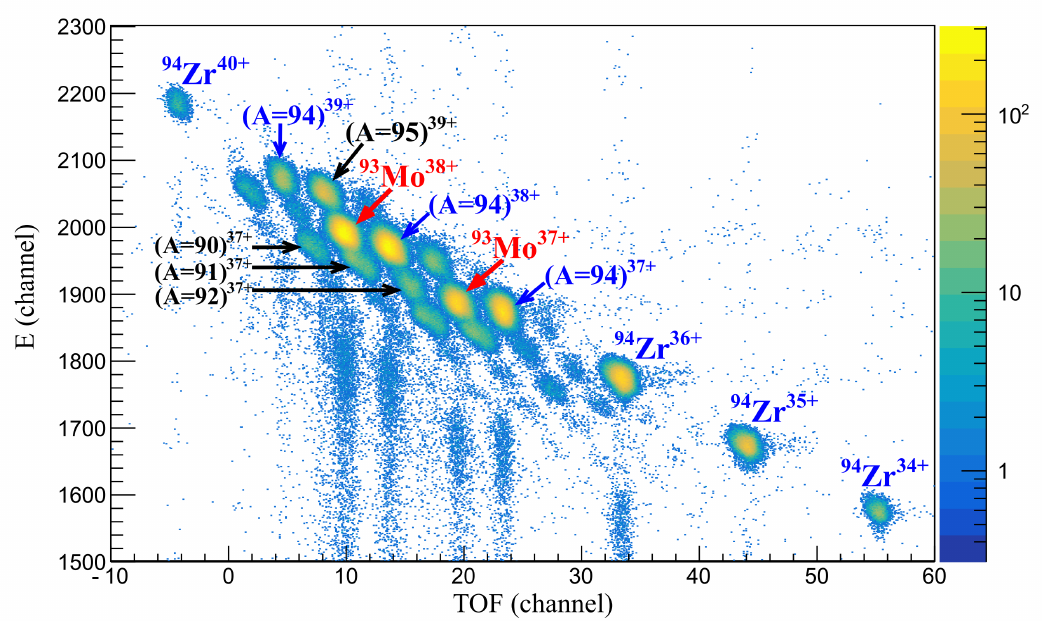}
	\caption{E vs TOF particle identification spectrum for production of $^{93m}$Mo. Reproduced with permission from Ref. \cite{Ma2026_93mMo_ChinaXiv}. The spectrum clearly resolves various isotopes and charge states produced in the $^{4}\mathrm{He}(^{94}\mathrm{Zr}, 5\mathrm{n})^{93m}\mathrm{Mo}$ reaction. The red arrows indicate the peaks corresponding to the $^{93}$Mo isotope in the 38+ and 37+ charge states, which were selected for subsequent measurements of Ref. \cite{ding2026isomer}.}
	\label{fig:tof_e_93Mo}
\end{figure}

To overcome these bottlenecks in energy and purity, the present cryogenic gas target system was employed to produce the beam via the $^{4}\mathrm{He}(^{94}\mathrm{Zr}, 5n)^{93m}\mathrm{Mo}$ reaction \cite{ding2026isomer}. By bombarding the cryogenic $^4$He gas target, maintained at approximately 980 mbar, with a 16.7 MeV/u $^{94}\mathrm{Zr}$ primary beam, the recoil energy of the isomeric products was increased to $\sim$11.6 MeV/u \cite{ding2026isomer}. Figure \ref{fig:tof_e_93Mo} displays a TOF-$E$ PID spectrum for production of $^{93m}$Mo \cite{ding2026isomer}. Consequently, the absolute beam intensity was increased by a factor of approximately 13.2 (reaching $\sim$5380 pps), with the in-flight purity significantly improved to nearly 20\%. By applying precise TOF gating in the offline analysis, the effective purity was further enhanced to approximately 50\% \cite{Ma2026_93mMo_ChinaXiv}.

This $^{93m}\mathrm{Mo}$ beam provided the prerequisite for directly investigating the Nuclear Excitation by Electron Capture (NEEC) mechanism in a low-background environment \cite{ding2026isomer}. Experimental measurements of Ref. \cite{ding2026isomer} indicate that during the slowing-down process in lead and carbon foils, the depletion probability of the isomer is in excellent agreement with theoretical predictions of inelastic nuclear scattering, contrary to the high-probability NEEC mechanism previously reported by the international academic community. This result resolved a decade-long controversy in nuclear isomer research, providing the most stringent upper limit on the NEEC depletion probability to date.

From an instrumentation perspective, the successful extraction of the intense and purified $^{93m}\mathrm{Mo}$ beam serves as a clear demonstration of the capabilities unlocked by the present cryogenic target.

\subsection{Thermal stability analysis}

Beam-induced heating is a critical issue for cryogenic gas targets operated with high-intensity primary beams. In the present work, the thermal behavior of the target was assessed through the deposited beam power, monitored cryogenic temperatures, and, for the $^{7}$Be production case, the observed linearity of the secondary-beam yield.

To quantify the thermal load on the target assembly, energy deposition was calculated using the SRIM code \cite{zigler010_SRIM}. Since the transmitted beam carries away most of the incident kinetic energy, relevant heat deposition occurs only via differential energy loss in the 2.5~$\mu$m front Havar window, the 80~mm gas volume, and the 2.5~$\mu$m back Havar window. Table~\ref{tab_thermal_deposition} summarizes the calculated thermal power deposition for all experimental conditions. For the most extreme condition (1.6~$\mu$A, 8.8~MeV/u $^7\mathrm{Li}^{3+}$), the maximum total deposited power estimated by SRIM was approximately 1.91 W, of which about 1.29 W was deposited in the hydrogen gas. For the 870 nA $^7\mathrm{Li}^{3+}$ and the $^{94}\mathrm{Zr}^{19+}$ beams, the estimated maximum total deposited powers were approximately 0.91 W ($\sim$0.57 W in hydrogen gas) and 1.48 W ($\sim$0.95 W in helium gas), respectively. These heat loads did not lead to observable operational instability of the LN$_2$-cooled target during the present measurements.

\begin{table*}
	\centering
	\caption{The maximum deposited power in the present target assembly estimated by SRIM \cite{zigler010_SRIM}.}
	\label{tab_thermal_deposition}
	\setlength{\tabcolsep}{4pt}
	\begin{tabular}{c l c c c c c c c }
		\toprule
		No. & Primary Beam & Energy & Current & Target Gas & $P_{\mathrm{front}}$ & $P_{\mathrm{gas}}$ & $P_{\mathrm{back}}$ & $P_{\mathrm{total}}$ \\ 
		& & (MeV/u) & (nA) &  & (W) & (W) & (W) & (W) \\ 
		\midrule
		1 & $^{15}\mathrm{N}^{7+}$ & 8.5 & 300 & $\mathrm{D}_2$ (530 mbar) & 0.13 & \textbf{0.34} & 0.14 & 0.62  \\
		2 & $^{15}\mathrm{N}^{7+}$ & 9.5 & 550 & $\mathrm{H}_2$ (800 mbar) & 0.23 & \textbf{0.88} & 0.24 & 1.34  \\
		3 & $^7\mathrm{Li}^{3+}$ & 8.8 & 1600 & $\mathrm{H}_2$ (1000 mbar) & 0.30 & \textbf{1.29} & 0.31 & 1.91  \\
		4 & $^7\mathrm{Li}^{3+}$ & 8.6 & 870 & $\mathrm{H}_2$ (800 mbar) & 0.17 & \textbf{0.57} & 0.17 & 0.91  \\
		5 & $^{94}\mathrm{Zr}^{19+}$ & 16.7 & 100 & $^4\mathrm{He}$ (980 mbar) & 0.25 & \textbf{0.95} & 0.27 & 1.48  \\
		\bottomrule
	\end{tabular}
\end{table*}

During beam irradiation, the thermocouple attached to the gas-cell outlet typically recorded temperatures of 82--86 K. While localized beam energy deposition possibly increases the gas temperature within the beam-interaction zone, a direct measurement of the beam-axis gas-temperature profile is beyond the scope of this commissioning work. Consequently, to avoid introducing model-dependent uncertainties, the target thicknesses quoted in Table~\ref{Tab_summary_present_RIB} are presented as nominal target thicknesses. These values are derived from the ideal-gas law using the monitored gas pressure, the 80~mm physical length, and the nominal 84 K operating temperature, serving as typical operational parameters rather than in-situ density profiles.

Despite the absence of direct local density profiling, the macroscopic secondary-beam yield exhibited an approximately linear dependence on the primary-beam intensity within the tested current range of $^7$Be beam, as shown in Fig. \ref{fig_7Be_intensity_linear_fit}. Together with the modest SRIM power-deposition estimates, this observation is consistent with the absence of large beam-induced thermal rarefaction under the present operating conditions.

\section{Discussion and Future Upgrades}

Compared to the prior alcohol-cooled target at RIBLL \cite{he2012}, the present $\mathrm{LN_2}$-cooled system demonstrates substantial advancements in target density, beam quality, and gas economy. By lowering the operating temperature from $\sim 2^\circ\mathrm{C}$ (275 K) to $\sim84$ K and increasing the operational pressure from $\sim 500$ mbar to 1000 mbar, the nominal target thickness has been enhanced by more than a factor of six. This densification elevates routine secondary beam intensities from $\sim 10^4$ pps levels to $10^5$--$10^6$ pps. Concurrently, the higher primary reaction yields synergize with the magnetic rigidity selection to improve the achievable purity for selected light RIBs, evolving from a typical $\sim 30\%$ to  85\%--90\% for $^7$Be, $>$99\% for $^{16}$N, and 95\% for $^{15}$O.

The advantage of the present $\mathrm{LN}_2$-cooled target is illustrated by the production of the heavy $^{93m}\mathrm{Mo}$ isomer beam. In the legacy alcohol-cooled system, the reported gas thickness ($\sim 0.7\ \mathrm{mg/cm}^2$) was dwarfed by the Havar window foils ($\sim 4.16\ \mathrm{mg/cm}^2$). This foil-dominated configuration caused severe multiple Coulomb scattering of the $^{94}\mathrm{Zr}$ primary beam, generating a massive kinematic background that completely obscured the rare secondary products. By increasing the nominal target thickness by a factor of more than six to $\sim 4.5 \mathrm{mg/cm}^2$, the $\mathrm{LN}_2$-cooled system transitions to a gas-thickness-comparable regime. This critical upgrade improves the reaction-yield-to-window-background ratio, allowing the better selection of $^{93m}\mathrm{Mo}$.

Furthermore, the underlying thermal management strategy has been optimized. The legacy system relied on a high-velocity forced gas circulation (without recycling) of 12.5 L/min to dissipate beam heat, resulting in large gas consumption. Conversely, under the present RIBLL beam intensities, stable operation was achieved without high-velocity forced gas circulation. The approximately 100--150 mL/min micro-flow was sufficient for pressure stabilization, while the heat load remained modest, as indicated by the SRIM power-deposition estimates.

Although the present cryogenic gas target exhibits excellent stability at RIBLL, its future application at the High Intensity heavy ion Accelerator Facility (HIAF) \cite{yang2013hiaf} poses higher thermal loads. HIAF is designed to deliver primary beam intensities up to 11 $\mu$A \cite{yang2023hiaf}, generating approximately 10--20 W estimated power deposition in gas target for light-element RIB. To accommodate such intense heat loads, the current system will be upgraded with reference to the mature forced-circulation designs established at CRIB \cite{yamaguchi2008}.

Specifically, the existing micro-flow venting configuration will be replaced by a closed-loop gas circulation system. This upgrade is expected to enhance convective heat removal while minimizing the consumption of expensive isotopic gases (e.g., $^3$He). Additionally, an automatic $\mathrm{LN_2}$ replenishment system will be integrated to support the uninterrupted, week-long operations typical of HIAF experimental campaigns. Concurrently, an in-line trace oxygen monitoring and interlock system will be integrated into the closed gas circulation loop to prevent hazardous explosive mixtures. These upgrades are expected to improve heat removal and target-density stability for future operation under primary-beam currents approaching 11~$\mu$A, with the goal of meeting the requirements of future operation for high-precision nuclear astrophysics and structure experiments.

\section{Summary}

In summary, a LN$_2$-cooled cryogenic gas target system was commissioned at RIBLL for in-flight RIB production. The target was operated with light gases at cryogenic temperatures and pressures up to 1000 mbar. The system was used to produce several low- and medium-energy secondary beams, including $^7\mathrm{Be}$, $^{16}\mathrm{N}$, $^{15}\mathrm{O}$, and  $^{93\mathrm{m}}\mathrm{Mo}$. Particularly, the high-spin isomer $^{93\mathrm{m}}\mathrm{Mo}$ beam provided the prerequisite for clarifying the long-standing NEEC controversy. The measured yields, PID spectra, and SRIM power-deposition estimates indicate stable operation under the tested beam conditions and are consistent with the absence of large beam-induced thermal rarefaction. 

These results establish the present LN$_2$-cooled gas target as an operating configuration for selected RIB production at RIBLL. Further work will be required to determine the local beam-axis gas density under higher beam-power conditions and to validate the planned closed-loop circulation scheme for future high-intensity operation.

\section{Acknowledgments}

The authors are grateful to the staff of HIRFL and RIBLL for the stable operation of the accelerators and their invaluable assistance. We extend our sincere gratitude to Prof. Jianjun He (Fudan University) for his insightful discussions. We are particularly indebted to Prof. Hidetoshi Yamaguchi (University of Tokyo) for his constructive suggestions and for sharing critical technical details regarding the design of the cryogenic gas target. This work was supported by the National Natural Science Foundation of China (11875329, 12335009, 12322509, U1632142),  GuangDong Basic and Applied Basic Research Foundation (2026A1515011317), National Key Research and Development program (MOST 2022YFA1602304).





\bibliographystyle{elsarticle-num}
\bibliography{LN2_gas_target_ref}







\end{document}